\documentclass[11pt]{article}
\usepackage{DGfest,epsfig}
\usepackage{times}

\begin{document}
\baselineskip 11.5pt
\title{GAUGE SYMMETRY OF THE THIRD KIND AND QUANTUM MECHANICS \\ 
       AS AN INFRARED PHENOMENON}
\author{H.-T. ELZE}

\address{Dipartimento di Fisica, Universit\`a di Pisa,\\ Largo Pontecorvo 3,
         I--56127 Pisa, Italia}

\maketitle\abstracts{We introduce functional degrees of freedom  
by a new gauge principle related to the phase of 
the wave functional. Thereby, quantum mechanical systems are seen as  
dissipatively embedded part of a nonlinear classical structure 
producing universal correlations. There are a fundamental length and 
an entropy/area parameter, besides standard couplings.  
For states that are sufficiently spread over 
configuration space, quantum field theory is recovered.}

\noindent
{\it It is a great pleasure to participate in the celebration 
and an honour to contribute this article to the Festschrift for Adriano Di Giacomo 
on occasion of his 70th birthday: With my very best wishes!}  

\section{Introduction} 
We consider ${\cal U}$(1) gauge transformations ``of the third kind'' which are 
restricted only to be local in the space of field 
configurations underlying quantum field theory (QFT). 

Our point of departure is the common observation 
that the potentiality described by the wave function(al) $\Psi$ and  
the reduction to the actuality of the outcome of any measurement process have been left 
outside of standard quantum theory: \\ 
{\it `` ... , it is an incomplete representation of 
real things, although it is the only one which can be built out of the fundamental 
concepts of force and material points (quantum corrections to classical mechanics). 
The incompleteness of the representation leads necessarily to the statistical 
nature (incompleteness) of the laws.''}~\cite{Einstein} \\ 
It appears that aspects of the theory that concern the concept of 
information have hitherto been left separate from 
the concepts of force and material points, the ``real things'' Einstein refers to. 
This is reflected, on the other side, in the remarkable derivation of the 
kinematical setting of quantum theory from three information theoretical constraints,    
as discussed, for example, in the review by Bub~\cite{Bub}: the  
``real things'' do not play a role in it.
  
One cannot help to feel that the cut between ``real things'' and ``information about 
real things'' is due to historical contingencies, not unlike the cut between dynamical 
theories, describing the effects of gravity in particular, and geometrical theories of 
space and time, until their fusion in general relativity.     

Reconsidering the role of the wave functional, our gauge principle is based on the 
assumption that its phase is not only subject to global gauge transformations (``of the first 
kind'') and local ones (``of the second kind'') related to its variables, which 
are the common fields of QFT. We introduce another class of  
functional gauge transformations which are {\it local in the space of field configurations},   
attributing a physical `charge' to $\Psi$.   
 
Guided by the gauge principle, 
we couple a dynamical functional ${\cal A}$ to the wave functional $\Psi$. 
We generalize the action of 
Dirac's variational principle, as applied to QFT, which is  
useful to study the interplay with the usual symmetries. Here,   
the would-be-quantum sector described by $\Psi$  
forms a dissipative subsystem of the enlarged structure, effectively 
extending it nonlinearly and nonlocally in the space of field configurations. 

We recall that much work has been done on nonlinear extensions of nonrelativistic 
quantum mechanics, most notably by Bialynicki-Birula and Mycielski~\cite{Iwo76} 
and by Weinberg~\cite{Weinberg89}. -- Works of Kibble~\cite{Kibble} and of Kibble and 
Randjbar-Daemi~\cite{Kibble80} present nonlinear extensions of 
scalar QFT models, the latter coupled to gravity, where parameters 
of the models are quantum state dependent. While this kind of nonlinearity is not compatible 
with our gauge principle, common feature is, of course, the effective {\it nonlinearity} (in 
$\Psi$) of the wave functional equation.

The local U(1) gauge invariance of the Schr\"odinger equation in quantum mechanics (``first quantization'') 
leads to the electromagnetic interaction of charged particles via the classical minimal coupling prescription. 
Here, we explore an analogous dynamical scheme for the functional Schr\"odinger equation 
(``second quantization''). 
It predicts the universal coupling of {\it all} fields that are     
variables in this equation and necessitates the introduction of a fundamental length. 

In this way, we obtain an embedding of  
quantum theory into an apparently classical framework, where the potentiality represented 
by $\Psi$ bears the character of a charge. It causes correlations of the underlying fields beyond what is encoded in their usual Lagrangians, with QFT describing an infrared limit. It is 
tempting to speculate that a model of this kind might point in the right direction towards putting the concepts of ``real things'' and ``information about real things'' on a common footing. 

In distinction to recent  
attempts to reconstruct quantum mechanics as an emergent 
theory~\cite{tHdet88,tHdet97,tHdet01,All,E04,I05,Smolin,Adler}, 
the present approach, so far, does not  
depend on a particular field theory, such as the Standard Model. 
We share, however, the tentative conclusion that quantum theory can and should be 
reconstructed as an effective theory describing large-scale  
behavior of {\it fundamentally deterministic degrees of freedom}~\cite{tHdet06}. 
Quantum states are no longer the primary degrees of freedom. 
Bell's theorem and the predicament 
of local hidden variable theories are circumvented, 
since the implicit nonlocality operates only at the pre-quantum level.
  
In Section 2, we recall the minimally coupled Schr\"odinger equation, 
in order to contrast it with the functional case in what follows. 
We stress that it amounts to introducing the nonrelativistic limit of a {\it classical} 
charged scalar field theory incorporating the electromagnetic interaction.   
This is relevant for the interpretation of our approach, where  
the wave functional $\Psi$ itself carries the new universal 
${\cal U}$(1) `charge' 
that acts as a source of the corresponding functional `field' ${\cal A}$. When 
the latter becomes negligible, ordinary QFT is recovered. 
Formal aspects are developed in Section 3 for the case of 
a single scalar field, while the extension incorporating other 
matter and gauge fields is briefly indicated and will be reported 
in detail elsewhere. We conclude with 
a list of problems and speculations left for further study.     
   
\section{Minimal Coupling in the Schr\"odinger Equation}
Textbooks expand on the topic of this section; for example, see Ref.~\cite{Saxon}, 
which contains clear discussions of the   
foundations of quantum mechanics and of its interpretation. We
represent it here, in order to distinguish the new 
element of the following section.  
   
We consider the single-particle Schr\"odinger equation in 
the coordinate representation: 
\begin{equation}\label{nrSchroedinger}
i\partial_t\psi (\vec x,t)=H\big (
\vec p\rightarrow\frac{1}{i}\nabla ,\vec x\big )\psi (\vec x,t) 
=\big\{ -\frac{1}{2m}\nabla^2+V(\vec x)\big\} \psi (\vec x,t)
\;\;,  \end{equation}
with the canonical momentum in the Hamiltonian $H$  
replaced by the derivative operator, $V$ representing an  
external potential, and $m$ denoting the particle mass, as usual. 
Units are chosen such that $\hbar =c=1$. 
   
The equation  
is invariant under {\it global} gauge transformations (``of the first kind'') of the 
form $\psi '=\exp (-i\lambda )\psi$, $\lambda$ real.  
Correspondingly, there is the continuity equation: 
\begin{equation}\label{probconserv}   
\partial_t\big (\psi^*\psi\big )+\nabla\cdot\big (
\psi^*\nabla \psi -\psi\nabla \psi^* \big )/2mi=0
\;\;, \end{equation}  
which expresses the local {\it probability conservation}, according    
to the probability amplitude interpretation 
of the wave function.  
 
Next, we consider {\it local} gauge transformations (``of the second kind''): 
\begin{equation}\label{localgt}
\psi '(\vec x,t)=\exp\left (-ie\lambda (\vec x,t)\right )\psi (\vec x,t)
\;\;, \end{equation}
with the coupling constant $e$ and $\lambda$ a real function.   
The Schr\"odinger equation (\ref{nrSchroedinger}) remains invariant,  
provided all derivatives are replaced by covariant derivatives, 
\begin{equation}\label{covderiv} 
\partial_\mu\equiv (\partial_t,\nabla )\rightarrow D_\mu\equiv 
\partial_\mu +ieA_\mu\equiv (\partial_t+ieA^0,\nabla -ie\vec A)
\;\;, \end{equation} 
and provided the vector potential transforms as 
\begin{equation}\label{vecAgt} 
A_\mu '(x)=A_\mu (x) +\partial_\mu\lambda (x)
\;\;, \end{equation} 
with $x^\mu\equiv (t,\vec x)$, considering Minkowski space  
with metric $g_{\mu\nu}\equiv\mbox{diag}(1,-1,-1,-1)$. The vector potential describes 
the electromagnetic field, $F_{\mu\nu}\equiv
\partial_\mu A_\nu -\partial_\nu A_\mu$, which is classical here. 
The related {\it current conservation} law is obtained from 
Eq.\,(\ref{probconserv}) using the substitution (\ref{covderiv}) and 
an overall multiplication by $e$.     
 
While all of this is familiar, two important remarks are in order: \\ 
\phantom .\hskip 0.15cm ({\bf A}) 
The quantum mechanical model of a charged particle interacting with 
the electromagnetic field descends from the  
classical Maxwell theory, via the minimally coupled classical particle Hamiltonian, 
through its quantization, 
finally to the gauge invariant Schr\"odinger equation. The point to 
make is that there is a {\it classical} regime where the quantum 
theory is anchored, in agreement with the Copenhagen interpretation~\cite{Saxon}. \\ 
\phantom .\hskip 0.15cm ({\bf B}) 
The free Schr\"odinger equation presents   
the nonrelativistic limit of the Klein-Gordon equation. The latter 
is not acceptable as   
a quantum mechanical single-particle wave equation, since there would be   
negative energy states and a `probability' density which is not positive definite~\cite{Saxon}. 
However, with minimal coupling, the Klein-Gordon 
equation describes   
{\it interacting classical} (!) {\it scalar and electromagnetic fields}.   
(Of course, quantization then leads to ``second quantized'' scalar electrodynamics.) 
 
In the following section, we show that QFT in the functional Schr\"odinger picture, 
even with all known fields and interactions included, 
can be extended in analogy to the Klein-Gordon equation by incorporating  
a functional ${\cal U}$(1) gauge symmetry (``of the third kind''). 
This may bring it close to a deterministic pre-quantum theory.           

\section{The Gauge Invariant Functional Wave Equation}
The functional Schr\"odinger 
picture of QFT is best suited for our argument. It is  
intuitively appealing by the resemblance to the case of first quantized  
systems. We refer to Refs.~\cite{FJ,JackiwRev,Kiefer,KieferWipf} 
for reviews, several applications, and further references. 

\subsection{The Scalar Field Case}  
Beginning with a generic scalar field theory, the functional Schr\"odinger 
equation is: 
\begin{equation}\label{fSscalar}
i\partial_t\Psi [\varphi ;t]=H[\hat\pi ,\varphi]\Psi [\varphi ;t]\equiv
\int\mbox{d}^3x\Big\{ -\frac{1}{2}\frac{\delta^2}{\delta\varphi^2}
+\frac{1}{2}(\nabla\varphi )^2+V(\varphi )\Big\}\Psi [\varphi ;t]
\;\;, \end{equation} 
where the Hamiltonian corresponds to the classical Hamiltonian density, 
including mass and selfinteraction terms in $V(\varphi )$. 
Implementing the quantization, the classical canonical momentum 
conjugate to the field (coordinate) $\varphi$ has been substituted according to: 
\begin{equation}\label{momentum}
\pi (\vec x)\;\longrightarrow\;\hat\pi (\vec x)\equiv
\frac{1}{i}\frac{\delta}{\delta\varphi (\vec x)}
\;\;, \end{equation}
i.e., in this coordinate representation,   
$\left [\varphi (\vec x),\hat\pi (\vec y)\right ]=i
\delta^3(\vec x-\vec y)\;$, as it should be.

In close analogy to gauge transformations in the 
first quantized Schr\"odinger picture, we now introduce  
the ${\cal U}$(1) {\it gauge transformation of the third kind}:  
\begin{equation}\label{localfuncgt} 
\Psi'[\varphi ;t]=\exp (-if\Lambda [\varphi ;t])\Psi [\varphi ;t]
\;\;, \end{equation} 
where $\Lambda$ denotes a time dependent real functional and $f$ a new 
dimensionless coupling constant. 
Note that this gauge transformation is local 
in the space of field configurations. The differerence from the usual    
gauge transformations in QFT will show up in the way we 
introduce covariant derivatives (see also Section 3.3). 

In fact, the wave functional equation (\ref{fSscalar}) 
can be made invariant under the transformation (\ref{localfuncgt}) by  
replacing derivatives by covariant ones:  
\begin{eqnarray}\label{dtcov}
\partial_t&\longrightarrow &{\cal D}_t\equiv
\partial_t+if{\cal A}_t[\varphi ;t]
\;\;, \\ \label{dxcov} 
\frac{\delta}{\delta\varphi (\vec x)}&\longrightarrow &
{\cal D}_{\varphi (\vec x)}\equiv
\frac{\delta}{\delta\varphi (\vec x)}+
if{\cal A}_\varphi [\varphi ;t,\vec x]
\;\;. \end{eqnarray} 
The real functional ${\cal A}$ presents the `potential' or `connection',   
analogous to the vector potential 
in Eqs.\,(\ref{covderiv}). Generally, ${\cal A}$ depends on $t$. However, 
it is a {\it functional} of $\varphi$ in Eq.\,(\ref{dtcov}), while it is a 
{\it functional field} in Eq.\,(\ref{dxcov}); 
the latter includes fields as a special case, for example, 
${\cal A}_\varphi [\varphi ;t,\vec x]={\cal A}(\varphi (\vec x);t)$. 
We distinguish these different components of ${\cal A}$ by the subscripts. 
Furthermore, the `potentials' need to transform according to: 
\begin{eqnarray}\label{Afunctionalgt} 
{\cal A}'_t[\varphi ;t]&=&{\cal A}_t[\varphi ;t]+
\partial_t\Lambda [\varphi ;t]
\;\;, \\ \label{Afunctiongt}  
{\cal A}'_\varphi [\varphi ;t,\vec x]&=&
{\cal A}_\varphi [\varphi ;t,\vec x]+
\frac{\delta}{\delta\varphi (\vec x)}\Lambda [\varphi ;t]
\;\;, \end{eqnarray} 
under the gauge transformation. Then, we may also define a `field strength':
\begin{equation}\label{field} 
{\cal F}_{t\varphi}[\varphi ;t,\vec x]\equiv 
\partial_t{\cal A}_\varphi [\varphi ;t,\vec x]
-\frac{\delta}{\delta\varphi (\vec x)}
{\cal A}_t[\varphi ;t]
\;\;, \end{equation} 
which is invariant under the transformations (\ref{Afunctionalgt}),\,(\ref{Afunctiongt}); 
note that ${\cal F}_{t\varphi}=[{\cal D}_t,{\cal D}_\varphi]/if$. 
     
In order to give a meaning to the coupling constant $f$, 
we have to postulate a consistent dynamics for the gauge potential 
${\cal A}$. 
All elementary fields supposedly are   
present as the coordinates on which the wave 
functional depends -- just a scalar field, besides time, in this 
section. Then, we may consider the following ${\cal U}$(1) invariant action:  
\begin{equation}\label{Action}
\Gamma\equiv\int\mbox{d}t\mbox{D}\varphi\;\Big\{ 
\Psi^*\Big ({\cal N}(\rho )
\stackrel{\leftrightarrow}{i{\cal D}}_t
-H[\frac{1}{i}{\cal D}_{\varphi},\varphi ]\Big )\Psi 
-\frac{l^2}{2}\int\mbox{d}^3x\;\big ({\cal F}_{t\varphi}
\big )^2\Big\}
\;\;, \end{equation} 
where    
$\Psi^*{\cal N}\stackrel{\leftrightarrow}{i{\cal D}}_t\Psi
\equiv\frac{1}{2}{\cal N}
\{\Psi^*i{\cal D}_t\Psi
+(i{\cal D}_t\Psi )^*\Psi\}$, and with   
a dimensionless real function ${\cal N}$ depending on the density:  
\begin{equation}\label{rho}
\rho [\varphi ;t]\equiv\Psi^*[\varphi ;t]\Psi [\varphi ;t]
\;\;. \end{equation}
We shall see shortly that ${\cal N}$ incorporates a necessary {\it nonlinearity}. 
The fundamental parameter $l$ has dimension 
$[l]=[length]$, for dimensionless measure $\mbox{D}\varphi$ and $\Psi$. 

Our action generalizes the 
action for the wave functional of a scalar field, which has been 
employed for applications of Dirac's 
variational principle to QFT, e.g., in Refs.~\cite{Kibble80,JackiwKerman}. 
The quadratic part in ${\cal F}_{t\varphi}$ is the 
simplest possible extension, i.e. local in  
$\varphi$ and quadratic in the derivatives, 
besides the newly introduced nonlinearity in $\rho$.  
-- 
The action depends on 
$\Psi ,\Psi^*,{\cal A}_t$, and 
${\cal A}_\varphi$ separately. 
While a Hamiltonian formulation is possible, we 
will obtain the equations of motion and a constraint directly by 
varying $\Gamma$ with respect to the variables. 
We assume that all necessary (functional) partial integrations are justified. 

Thus, varying $\Gamma$ with respect to $\Psi^*$ (and $\Psi$)
yields the gauge invariant $\Psi$-functional equation of motion (and its adjoint):  
\begin{equation}\label{ginvfSscalar}
\left (\rho {\cal N}(\rho )\right )' 
i{\cal D}_t\Psi [\varphi ;t]
=H[\frac{1}{i}{\cal D}_{\varphi},\varphi ]
\Psi [\varphi ;t]
\;\;, \end{equation}
replacing the Schr\"odinger equation (\ref{fSscalar}); 
here $f'(\rho )\equiv\mbox{d}f(\rho )/\mbox{d}\rho$. 
Varying with respect to ${\cal A}_\varphi$,  
we obtain the invariant `gauge field equation': 
\begin{equation}\label{fieldeq} 
\partial_t{\cal F}_{t\varphi}[\varphi;t,\vec x]
=\frac{f}{2il^2}\left ( 
\Psi^*[\varphi;t]{\cal D}_{\varphi (\vec x)}\Psi [\varphi;t]
-\Psi [\varphi;t]({\cal D}_{\varphi (\vec x)}\Psi [\varphi;t])^*
\right )
\;\;, \end{equation}
which completes the set of dynamical equations. 

The nonlinear Eq.\,(\ref{ginvfSscalar}) preserves the normalization, i.e. 
any imposed value of  
$\langle\Psi |\Psi\rangle\equiv\int\mbox{D}\varphi\;\Psi^*\Psi$ is conserved,  
while the overlap of two different states, $\langle\Psi_1|\Psi_2\rangle$, 
generally varies in time. While this may hint at a probability interpretation, 
the continuity equation, Eq.\,(\ref{continuity}) below, shows that this cannot  
be maintained, in general. For ${\cal A}\neq 0$, also the 
{\it homogeneity property} does no longer hold, 
i.e., $\Psi$ and $z\Psi$ ($z\in\mathbf{Z}$) present two different  
physical states~\cite{Iwo76,Weinberg89}. 
This changes an essential aspect of the measurement theory~\cite{Kibble} and    
indicates that here QFT is embedded in a {\it classical} framework (see Section 3.2).   
  
Furthermore, note that the Hamiltonian $H$, 
unlike in QFT, 
cannot be arbitrarily shifted by a constant $\Delta E$, transforming  
$\Psi\rightarrow\exp (-i\Delta Et)\Psi$. Our action is, however, invariant under space-time translations and spatial rotations. The behavior under Lorentz boosts is more difficult 
to assess (cf. Ref.~\cite{Kibble}) 
and will be considered elsewhere. 
  
In the action there is no time derivative acting on the variable 
${\cal A}_t$ which, therefore, acts as a Lagrange multiplier for a constraint. 
Thus, variation with respect to ${\cal A}_t$ yields the  
corresponding gauge invariant `Gauss' law':
\begin{equation}\label{Gauss}
-\int\mbox{d}^3x\;\frac{\delta}{\delta\varphi (\vec x)}{\cal F}_{t\varphi}[\varphi ;t,\vec x]
=\frac{f}{l^2}\Psi^*[\varphi ;t]\Psi [\varphi ;t]{\cal N}(\rho )
\;\;. \end{equation} 
This can be combined with Eq.\,(\ref{fieldeq}) to result in the continuity equation: 
\begin{equation}\label{continuity}
0=\partial_t\Big (\rho{\cal N}(\rho )\Big )
+\frac{1}{2i}\int\mbox{d}^3x\;\frac{\delta}{\delta\varphi (\vec x)}
\left ( 
\Psi^*{\cal D}_{\varphi (\vec x)}\Psi 
-\Psi ({\cal D}_{\varphi (\vec x)}\Psi )^*
\right )
\;, \end{equation}
which expresses the local ${\cal U}$(1) `charge' conservation in the space of field configurations. 

Furthermore, 
functionally integrating Eq.\,(\ref{Gauss}), we find that the 
total `charge' $Q$ has to vanish at all times: 
\begin{equation}\label{charge}
Q(t)\equiv\frac{f}{l^2}\int\mbox{D}\varphi\;\rho{\cal N}(\rho )=0
\;\;, \end{equation}    
since the functional integral of a total derivative is zero. 
Here, the necessity of the nonlinearity becomes obvious. 
Without it, the vanishing total `charge' could not be implemented.  
However, besides necessarily multiplying the invariant term  
$\Psi^*i{\cal D}_t\Psi$ in the action, the nonlinearity is still undetermined. 

Mean field type nonlinearities have been considered in Refs.~\cite{Kibble,Kibble80} before, 
with nonlinear modifications of the underlying Hamiltonian $H$.  
They are based on a field $\bar\rho$:     
\begin{equation}\label{rhobar}
\bar\rho (\vec x)\equiv 
\frac{\int\mbox{D}\varphi\;
\Psi^*[\varphi ;t]\varphi^n(\vec x)\Psi [\varphi ;t]}
{\int\mbox{D}\varphi '\;\Psi^*[\varphi ';t]\Psi [\varphi ';t]}
\;\;, \end{equation} 
e.g., with $n=2$, which here is ${\cal U}$(1) invariant. Nevertheless, 
this type of nonlinearity leads to a conflict with the 
condition of vanishing total charge.  

Anticipating some justification by the following analysis, we presently choose: 
\begin{equation}\label{Nlog}
\rho{\cal N}(\rho )\equiv\rho\log (\rho )+\rho S
\;\;, \end{equation}
with $S$ a dimensionless parameter. We consider $\rho\log (\rho )$ as the negative of 
an {\it entropy density}, recalling that the density $\rho$ can be 
normalized, $\int\mbox{D}\varphi\;\rho =1$, independent of time. 
Furthermore, it should be replaced by a proper {\it probability} in this entropy term: 
\begin{equation}\label{prob} 
\rho_P[\varphi ]\equiv
\int_{\varphi -\Delta\varphi /2}^{\varphi +\Delta\varphi /2}\mbox{D}\varphi'\rho [\varphi']
\;\;, \end{equation} 
i.e., the probability to find the system within an interval $\Delta\varphi$ 
around the field configuration $\varphi$. The natural scale for $\Delta\varphi$ is $l^{-1}$.
If $\rho$ is essentially constant over such an interval, then the coarse-graining leading 
to $\rho_P$ simply amounts to muliplication by its volume, which needs to be regularized. 
We keep this in mind and work with Eq.\,(\ref{Nlog}).     

With this choice, the time derivative of $\Psi$ in Eq.\,(\ref{ginvfSscalar}) is 
multiplied by:   
\begin{equation}\label{prefactor}
\left (\rho {\cal N}(\rho )\right )'=1+S+\log (\rho )
\;\;. \end{equation} 
Thus, the timescale of the $\Psi$-functional evolution effectively shrinks or 
expands, depending on the probability to find the system in the respective region of 
configuration space. 

We conclude this section with several observations: \\  
\phantom .\hskip 0.15cm ({\bf A}) 
The dimensionless coupling constant $f$ can be absorbed by rescaling:  
\begin{equation}\label{rescaling}
{\cal A}\;\longrightarrow\; 
f^{-1}{\cal A}     
\;\;. \end{equation} 
The resulting factor $f^{-2}$ multiplying $l^2$ in the action simply  
redefines this parameter. Consequently, we set $f=1$ from now on. --  
Furthermore, our equations involve second functional derivatives at coinciding 
points which should be regularized. 
We do not perform this here and proceed heuristically. \\    
\phantom .\hskip 0.15cm ({\bf B}) 
The system of Eqs.\,(\ref{ginvfSscalar})--(\ref{Gauss}), 
and (\ref{Nlog}) obeys a 
{\it weak superposition principle}~\cite{Iwo76}: The sum of two solutions, $\Psi_{1,2}$, 
that do not overlap, presents also a solution, provided that 
${\cal A}={\cal A}_1+{\cal A}_2$ is determined consistently. \\  
\phantom .\hskip 0.15cm ({\bf C}) 
It seems questionable whether our nonlinear extension of QFT is {\it local} 
in the usual sense \cite{Kibble}: 
Leaving only $i\partial_t\Psi$ on the left-hand side of Eq.\,(\ref{ginvfSscalar}), the 
resulting evolution operator on the other side cannot be written as integral of a density 
$\tilde H(\vec x)$, in order to check whether   
$[\tilde H(\vec x),\tilde H(\vec x')]=0$, for $\vec x\neq\vec x'$, meaning locality. 
Yet, instead, the system may be perturbed by a small source added to the 
action, such as $\int\mbox{d}t\mbox{D}\varphi\;\Psi^*J[\varphi ;t]\Psi$. 
This does not modify the `gauge field equation' and `Gauss' law'.  
If the source is confined to a localized region $\triangle$ and acts only at $t=t_0$, 
$J[\varphi ;t]\equiv\delta (t-t_0)\int_{\triangle}\mbox{d}^3x\;j(\vec x)\varphi (\vec x)$,  
for example, 
then $\Psi [\varphi ;t_0+\mbox{d}t]$ shows no effect when probed by an operator with 
support $\{\vec x |\vec x\notin\triangle\}$. As usual, only the derivative term 
${\cal D}_\varphi^{\;2}$ in $H$ spreads the perturbation ({\it microcausality}).  
-- However, suppose we integrated out the `gauge field'.  
The resulting effective equation for $\Psi$ would  
be {\it nonlocal in field space and in space-time}. \\  
\phantom .\hskip 0.15cm ({\bf D}) 
We find from Eqs.\,(\ref{Gauss}), (\ref{Nlog}) the interesting result 
that the `charge' density $\rho{\cal N}(\rho )$ 
is the deviation of entropy density per unit area  
from the reference density $\rho S/l^2$. 
The total entropy is constrained to equal $S$ by Eq.\,(\ref{charge}). Therefore,  
also the {\it entropy/area} $S/l^2$  
is a parameter here, besides the fundamental {\it length} $l$. -- 
We remark that entropy per area is an essential parameter in  
seemingly unrelated work of Padmanabhan~\cite{Padmanabhan}, where it is suggested that 
gravity is intrinsically holographic and quantum mechanical.   

\subsection{Stationary States, Separability and QFT Limit}
We study the separation of the time dependence in the coupled 
Eqs.\,(\ref{ginvfSscalar})--(\ref{Gauss}), assuming:  
\begin{equation}\label{oscill}
\Psi [\varphi ;t]\equiv e^{-i\omega t}\Psi_\omega [\varphi ]
\;\;, \end{equation} 
with $\omega\in\mathbf{R}$, and consistently 
assume {\it time independent} ${\cal A}$-functionals. With the nonlinearity 
of Eq.\,(\ref{Nlog}), we obtain from Eq.\,(\ref{ginvfSscalar}): 
\begin{equation}\label{PsiZeroEq} 
\Big (1+S+\log (\rho_\omega )\Big )
\Big (\omega -{\cal A}_t[\varphi ]\Big )\Psi_\omega [\varphi ]
=H[\frac{1}{i}{\cal D}_{\varphi},\varphi ]
\Psi_\omega [\varphi ]
\;\;, \end{equation}
with ${\cal D}_\varphi = \frac{\delta}{\delta\varphi}+i{\cal A}_\varphi$ and  
$\rho_\omega\equiv\Psi_\omega^*[\varphi ]\Psi_\omega [\varphi ]$. 
From Eq.\,(\ref{fieldeq}) follows: 
\begin{equation}\label{fieldeq0} 
\frac{1}{2i}\left ( 
\Psi^*_\omega [\varphi ]{\cal D}_{\varphi (\vec x)}\Psi_\omega [\varphi ]
-\Psi_\omega [\varphi ]({\cal D}_{\varphi (\vec x)}\Psi_\omega [\varphi ])^*
\right )=0
\;\;, \end{equation}
which expresses the vanishing of the `current' in the stationary 
situation. 

Applying a time independent gauge transformation, 
cf. Eqs.\,(\ref{localfuncgt}), (\ref{Afunctiongt}), the stationary wave functional 
can be made {\it real}. The Eq.\,(\ref{fieldeq0}) then implies ${\cal A}_\varphi=0$;   
consequently, we replace ${\cal D}_{\varphi}\rightarrow\frac{\delta}{\delta\varphi}$ 
everywhere. Finally, `Gauss' law', Eq.\,(\ref{Gauss}), 
determines ${\cal A}_t$: 
\begin{equation}\label{Gauss0}  
\int\mbox{d}^3x\;\frac{\delta^2}{\delta\varphi (\vec x)^2}{\cal A}_t[\varphi ]
=\frac{1}{l^2}\rho_\omega\Big (S+\log (\rho_\omega )\Big )
\;\;, \end{equation} 
incorporating Eq.\,(\ref{Nlog}). 
This has to be solved selfconsistently together with Eq.\,(\ref{PsiZeroEq}). 

Separation of the time dependence thus leads to two coupled 
equations. One may guess an appropriate time independent 
$\Psi_\omega$-functional, for example some kind of Gaussian. 
Having an action at hand, Eq.\,(\ref{Action}), the parameters of such an Ansatz can be optimized according to the variational principle. This is analogous      
to Hartree approximation and semiclassical limit of QFT~\cite{Kiefer,JackiwKerman}. 
Furthermore, the Eq.\,(\ref{Gauss0}) can be solved, at least formally, by functional Fourier 
transformation. In this way, the `connection' ${\cal A}_t$ can be eliminated, at the 
expense of introducing the nonlocality mentioned before. 

Let us turn to the question of {\it separability}. This is an important property of linear 
quantum theory. It allows to combine subsystems which do not interact with each other, without 
creating unphysical correlations between them~\cite{Iwo76,Weinberg89}. This should be preserved for the phenomena and to the level of accuracy 
which are accessible experimentally. 

We will give a heuristic argument that linear {\it QFT 
arises in the infrared (IR) limit}, and consequently separability. Discretizing 
the system on a finite spatial lattice, would be a first step towards a more 
rigorous derivation.  
  
Considering the stationary equations, let us assume that the system is in a  
{\it diffuse state}, characterized by a density $\rho_\omega$ that is widely spread 
over the space of fields $\varphi$. In such a high entropy state, 
we have that $\rho_\omega$ or, rather, the corresponding probability $\rho_P[\varphi ]$ 
of Eq.\,(\ref{prob}) is nearly homogeneous and small everywhere. Introducing the  
density of the Hamiltonian of Eq.\,(\ref{PsiZeroEq}), via $H\equiv\int\mbox{d}^3x\;H(\vec x)$, 
the local energy density associated with such a (real) state $\Psi_\omega$, i.e. 
$\epsilon (\vec x)\equiv\int\mbox{D}\varphi\;\Psi_\omega H(\vec x)\Psi_\omega$, must be 
small, in particular, since also the kinetic term $\propto\delta_\varphi^{\;2}\Psi_\omega$ 
cannot be too large (IR limit).  
  
In this situation, assuming $\rho_\omega\approx\exp (-S)$, the `charge density' on the 
right-hand side of Eq.\,(\ref{Gauss0}) is very small. 
Therefore, we neglect ${\cal A}_t$ compared to $\omega$ in 
Eq.\,(\ref{PsiZeroEq}). Furthermore, the nonlinear prefactor 
there goes to one in the same limit.    
Thus, the stationary functional Schr\"odinger equation results: 
\begin{equation}\label{approxS}
\omega\Psi_\omega =H\Psi_\omega 
\;\;, \end{equation} 
and with it the known structures of QFT. -- 
It will be most interesting to explore consequences of the 
small violations of the linear equation (\ref{approxS}), due to terms which 
involve $\rho_\omega$ or ${\cal A}_t$. Similarly as in the mean field type model 
of Kibble~\cite{Kibble}, nonlinear effects become important in our framework only for 
states with a small uncertainty in configuration space, such that the IR limit 
does not apply. 

Finally, we remark that two stationary solutions, $\Psi_{\omega_{1,2}}$, of the present 
eigenvalue problem, in general, obey an {\it orthogonality} relation: 
\begin{eqnarray}
&\;&0=\int{\cal D}\varphi\;\Psi_1^*[\varphi ]\Psi_2[\varphi ]
\Big ( 
(1+S+\log (\rho_{\omega ,1}))(\omega_1-{\cal A}_{t1})
\nonumber \\ [1ex] \label{orthogonality}
&\;&\;\;\;\;\;\;\;\;\;\;\;\;\;\;\;-
(1+S+\log (\rho_{\omega ,2}))(\omega_2-{\cal A}_{t2})\Big )
\;\;, \end{eqnarray} 
where ${\cal A}_{t\;1,2}$ are determined selfconsistently 
by Eq.\,(\ref{Gauss0}). This reduces to the usual one of QFT 
in the IR limit just discussed.  

\subsection{Internal Symmetries}
Some modifications are necessary when the scalar field model  
of Section 3.1 is replaced by a more realistic field theory of 
interacting matter and gauge fields. For simplicity, we restrict ourselves 
to the Abelian Higgs model or scalar QED \cite{Kiefer}. The  
formalism to treat fermions in the Schr\"odinger picture with the help 
of functionals depending on Grassmann variables can be found, for example, 
in Refs.\,\cite{FJ,KieferWipf}. 
  
Working in the temporal axial gauge for the electromagnetic gauge potential, 
$A^0=0$, we begin with the classical Hamiltonian: 
\begin{equation}\label{HamiltonianQED} 
H[\pi ,\pi^*,\vec\Pi ,\Phi ,\Phi^*,\vec A]
=\int\mbox{d}^3x\left\{\pi\pi^* +\vec D\Phi\cdot(\vec D\Phi )^*
+\frac{1}{2}\left (\vec\Pi^2+\vec B^2\right )+ 
V(\Phi\Phi^*)\right\}  
\;\;, \end{equation} 
describing the charged (complex) scalar field $\Phi$ in interaction with 
the gauge field $\vec A$ and a quadratic selfinteraction 
term $V$; the covariant derivative $\vec D_\mu$ is as in Eq.\,(\ref{covderiv}). 
There is the magnetic field contribution   
$\vec B^2=\frac{1}{2}F_{ij}F^{ij},\; i,j=1,2,3$, $F_{ij}\equiv\partial_iA_j-\partial_jA_i$, 
and the canonical momenta $\pi ,\pi^*,\vec\Pi$ are obtained from the corresponding 
Lagrangian. Since $A^0$ is not dynamical, the Hamiltonian has to be supplemented by  
Gauss' Law: 
\begin{equation}\label{GaussQED}
\nabla\cdot\vec\Pi =-ie\left (\Phi\pi -\Phi^*\pi^*\right )
\;\;. \end{equation} 
We recall the relation with the electric field, 
$\Pi^i =E_i\equiv F_{0i}\equiv\partial_0A_i$, in 
this gauge.  
  
The theory is quantized in the coordinate representation (cf. Section 3.1) by: 
\begin{eqnarray}\label{phiquant}
\pi^{(*)}&\longrightarrow&\hat\pi^{(\dagger )}\equiv\frac{1}{i}\frac{\delta}{\delta\Phi^{(*)}}
\;\;, \\ [1ex] \label{Aquant} 
\Pi_j&\longrightarrow&\hat\Pi_j\equiv-\frac{1}{i}\frac{\delta}{\delta A^{j}}  
\;\;, \end{eqnarray} 
which, in particular, implies the commutation relation: 
$[A^j(\vec x),\Pi_k(\vec y)]=i\delta^j_k\delta^3(\vec x-\vec y)$. Furthermore, then, the quantized version of Gauss' Law is obtained:   
\begin{equation}\label{GaussQED1}
\Big\{\partial^j\frac{\delta}{\delta A^j}
-ie\Big (\Phi\frac{\delta}{\delta\Phi}-\Phi^*\frac{\delta}{\delta\Phi^*}\Big )  
\Big\}\Psi [\Phi ,\Phi^*,\vec A;t]=0
\;\;. \end{equation} 
It expresses the invariance of the wave functional under U(1) gauge transformations:   
\begin{equation}\label{QEDtrans} 
\vec A'(\vec x)=\vec A(\vec x)-\nabla\lambda (\vec x)\;\;,\;\;\; 
\Phi '(\vec x)=\Phi (\vec x)\exp [-ie\lambda (\vec x)]
\;\;, \end{equation} 
with $\lambda$ an arbitrary real function. So far, this presents the 
starting point for a study of the Abelian Higgs model or scalar QED in the functional 
Schr\"odinger picture \cite{Kiefer}.   
   
The question must be raised now, whether the electromagnetic U(1) symmetry can {\it coexist}
with the new ${\cal U}$(1) symmetry, which we introduced in Section 3.1. The answer is 
positive, as we shall demonstrate in the following. 
  
As before, we replace the time derivative and the functional derivatives, which  
appear in the Hamiltonian or in Gauss' Law, by covariant ones: 
\begin{eqnarray}\label{dtcov1}
\partial_t&\longrightarrow &{\cal D}_t\equiv
\partial_t+i{\cal A}_t[\Phi ,\Phi^*,\vec A;t]
\;\;, \\ \label{dXicov} 
\frac{\delta}{\delta\Xi (\vec x)}&\longrightarrow &
{\cal D}_{\Xi (\vec x)}\equiv
\frac{\delta}{\delta\Xi (\vec x)}+
i{\cal A}_\Xi [\Phi ,\Phi^*,\vec A;t,\vec x]
\;\;, \end{eqnarray}  
where $\Xi$ represents any one of the fields $\Phi ,\Phi^*,\vec A$, on which the 
wave functional depends. For each one, we presently need a separate `gauge potential' 
${\cal A}_\Xi$, which is minimally coupled. This implements the covariance under 
the ${\cal U}$(1) transformations:
\begin{eqnarray}\label{psifunctionalgt1}  
\Psi'[\tilde\Xi;t]&=&
\exp (-i\Lambda [\tilde\Xi;t])\Psi [\tilde\Xi;t]
\;\;, \\ [1ex] \label{Afunctionalgt1} 
{\cal A}'_t[\tilde\Xi;t]&=&{\cal A}_t[\tilde\Xi;t]+
\partial_t\Lambda [\tilde\Xi;t]
\;\;, \\ [1ex] \label{Afunctiongt1}  
{\cal A}'_\Xi [\tilde\Xi;t,\vec x]&=&
{\cal A}_\Xi [\tilde\Xi;t,\vec x]+
\frac{\delta}{\delta\Xi (\vec x)}\Lambda [\tilde\Xi;t]
\;\;, \end{eqnarray} 
for the present model; we collectively denote $\Phi ,\Phi^*,\vec A$ by $\tilde\Xi$ 
from now on. As before, we 
define an invariant `electric field strength': 
\begin{equation}\label{field1} 
{\cal F}_{t\Xi}[\tilde\Xi;t,\vec x]\equiv 
\partial_t{\cal A}_\Xi [\tilde\Xi;t,\vec x]
-\frac{\delta}{\delta\Xi (\vec x)}
{\cal A}_t[\tilde\Xi;t]
\;\;, \end{equation} 
one for each field. We may also define an invariant bilocal `magnetic field strenght':  
\begin{equation}\label{field2} 
{\cal F}_{\Xi\Xi'}[\tilde\Xi;t,\vec x,\vec y]\equiv 
\frac{\delta}{\delta\Xi (\vec x)}{\cal A}_{\Xi'}[\tilde\Xi;t,\vec y]
-\frac{\delta}{\delta\Xi' (\vec y)}
{\cal A}_\Xi [\tilde\Xi;t,\vec x]
\;\;, \end{equation} 
which does not exist in the single scalar field case of Section 3.1.

Then, an ${\cal U}$(1) and U(1) invariant action for Abelian Higgs model or 
scalar QED is: 
\begin{eqnarray}
\Gamma &\equiv&\int\mbox{d} t\mbox{D}\Phi\mbox{D}\Phi^*\mbox{D}\vec A\;\Big\{ 
\Psi^*[\tilde\Xi;t]\Big ({\cal N}(\rho )
\stackrel{\leftrightarrow}{i{\cal D}}_t
-H[\frac{1}{i}{\cal D}_{\tilde\Xi},\tilde\Xi ]\Big )\Psi [\tilde\Xi;t]
\nonumber \\ [1ex] \label{Action1}
&\;&-\frac{l^2}{2}\sum_\Xi\int\mbox{d}^3x\;\left ({\cal F}_{t\Xi}\right )^2
-\frac{l'}{4}\sum_{\Xi,\Xi'}\int\mbox{d}^3x\mbox{d}^3y\;
\left ({\cal F}_{\Xi\Xi'}\right )^2
\Big\}
\;\;, \end{eqnarray} 
with Hamiltonian of Eq.\,(\ref{HamiltonianQED}), quantized according 
to Eqs.\,(\ref{phiquant})--(\ref{Aquant}), and with time and functional derivatives replaced 
by covariant ones, Eqs.\,(\ref{dtcov1})--(\ref{dXicov}). For dimensional reasons, 
the `magnetic field squared' contribution comes with a {\it length} parameter $l'$, which at 
this point can be chosen differently from $\sqrt{l^2}$. 
  
We do not discuss the equations of motion here, which simply generalize  
the ones of Section 3.1. However, turning to the `Gauss' Law' which is due to 
the `potential' ${\cal A}_t[\tilde\Xi;t]$ acting as a Lagrange multiplier, we 
presently obtain: 
\begin{equation}\label{Gauss1}
\sum_\Xi\int\mbox{d}^3x\;\frac{\delta}{\delta\Xi (\vec x)}{\cal F}_{t\Xi}[\tilde\Xi;t,\vec x]
=\frac{1}{l^2}\rho {\cal N}(\rho )
\;\;, \end{equation} 
with $\rho\equiv\Psi^*[\tilde\Xi ;t]\Psi [\tilde\Xi ;t]$.  
Thus, besides Eq.\,(\ref{GaussQED1}), we have a second constraint.  

The two constraints respectively express ${\cal U}(1)$ and U(1) gauge invariance.  
One way to arrange the coexistence of these symmetries consists 
in decoupling the respective constraints. --  
Having replaced the functional derivatives in Eq.\,(\ref{GaussQED1}) by the covariant 
ones from Eq.\,(\ref{dXicov}), let the ${\cal A}_\Xi$-functionals 
which enter be {\it pure gauge} with respect to ${\cal U}(1)$: 
\begin{equation}\label{puregauge} 
{\cal A}_\Xi [\tilde\Xi;t,\vec x]=
\frac{\delta}{\delta\Xi (\vec x)}\alpha [\tilde\Xi;t]
\;\;, \end{equation} 
with $\alpha$ an U(1) invariant functional. In this case, their contributions 
to the ordinary Gauss Law (\ref{GaussQED1}) cancel, which is thus kept intact,  
while being invariant under ${\cal U}(1)$. If we, furthermore, 
restrict  ${\cal A}_t[\tilde\Xi;t]$ to be U(1) invariant, then this is also true of  
the `Gauss Law' (\ref{Gauss1}), since the right-hand side 
is invariant already by Eq.\,(\ref{GaussQED1}). 
In this restricted situation, the `magnetic field' (\ref{field2}) vanishes. 
Correspondingly, the `magnetic' contribution to the action $\Gamma$, Eq.\,(\ref{Action1}), 
and the dependence on the length parameter $l'$ disappear. What is left is the minimally 
coupled ${\cal U}(1)$ dynamics, incorporating only ${\cal A}_t$, which does not 
interfere with the internal U(1) symmetry of the Abelian Higgs model or 
scalar QED. 

Nothing prevents the extension of this construction 
to more realistic models, incorporating non-Abelian gauge 
symmetries, for example. 

\section{Conclusions}
Gauge transformations ``of the third kind'' which attribute a ${\cal U}$(1) `charge'  
to the wave functional have been introduced here. This leads to an embedding of 
quantum field theory in a larger nonlinear structure, which differs from the 
earlier proposals of nonlinear generalizations of quantum mechanics or  
QFT~\cite{Iwo76,Weinberg89,Kibble,Kibble80}. 

We tentatively interpret it 
as a classical one, since real and imaginary part of the wave functional $\Psi$ are   
related to its differently charged components which, besides being governed by the interactions 
of the underlying field theory model, are coupled through a new connection functional 
${\cal A}$. When effects of the latter are negligible, QFT is recovered.     
  
A number of interesting problems clearly need further study, before this proposal 
can stand on its own: \begin{itemize} 
\item A theory of the observables and the measurement process needs to be worked out.  
It seems promising that the energy-momentum tensor following from our action, 
Eq.\,(\ref{Action}) or Eq.\,(\ref{Action1}), is the one of a quantized scalar or 
Abelian Higgs model (scalar QED), respectively, 
{\it plus} contributions due to the coupling between 
$\Psi$ and ${\cal A}$. When the latter is small, the usual observables might be useful, 
while the coupling might be important for the reduction or collapse of the wave 
functional. 
\item The symmetry properties of our extended theory need to be considered in  
detail, particularly in view of the introduction of the fundamental length 
$l$.\footnote{Under suitable assumptions about the background metric, the action $\Gamma$  
can be written in explicitly Lorentz invariant form, to be discussed elsewhere.}
\item A solution in the case of an underlying free field theory  
should be possible, based on the variational principle, for example. This will be helpful 
to better understand the new coupling and induced nonlocal correlations. 
\item Besides the indicated extension for coupled gauge and matter fields, 
reparametrization invariant models are an important target. As compared 
to a Wheeler-DeWitt type equation, giving rise to the ``problem of time'' and ensuing 
difficulties to understand evolution in a frozen picture, the presence of  
additional nonlinear terms in what replaces this equation might actually be useful.       
\item 
Is there more to the choice of the necessary nonlinearity ${\cal N}(\Psi^*\Psi )$, 
Eq.\,(\ref{Nlog}), and its relation to an entropy/area parameter~\cite{Padmanabhan} and 
information?
\end{itemize}


\end{document}